\documentclass[%
 aps,
 prl,
 twocolumn,
 superscriptaddress,
 amsmath,amssymb,
 nofootinbib,
 natbib,
]{revtex4-2}

\usepackage[hidelinks]{hyperref}
\usepackage{graphicx}
\usepackage{dcolumn}
\usepackage{bm}
\usepackage{booktabs}
\usepackage[english]{babel}

\newcommand{\aanda}{Astron.\ Astrophys.}
\newcommand{\apjl}{Astrophys.\ J.\ Lett.}
\newcommand{\mnras}{Mon.\ Not.\ R.\ Astron.\ Soc.}
\newcommand{\jcap}{J.\ Cosmol.\ Astropart.\ Phys.}

\begin{document}

\title{A Bayesian Perspective on Evidence for Evolving Dark Energy}

\author{Dily Duan Yi Ong}
\email{dlo26@cam.ac.uk}
\affiliation{Kavli Institute for Cosmology, University of Cambridge, Madingley Road, Cambridge, CB3 0HA, U.K.}
\affiliation{Cavendish Laboratory, University of Cambridge, J.J. Thomson Avenue, Cambridge, CB3 0HE, U.K.}
\affiliation{Institute of Astronomy, University of Cambridge, Madingley Road, Cambridge, CB3 0HA, U.K.}

\author{David Yallup}
\affiliation{Kavli Institute for Cosmology, University of Cambridge, Madingley Road, Cambridge, CB3 0HA, U.K.}
\affiliation{Institute of Astronomy, University of Cambridge, Madingley Road, Cambridge, CB3 0HA, U.K.}

\author{Will Handley}
\affiliation{Kavli Institute for Cosmology, University of Cambridge, Madingley Road, Cambridge, CB3 0HA, U.K.}
\affiliation{Institute of Astronomy, University of Cambridge, Madingley Road, Cambridge, CB3 0HA, U.K.}

\date{\today}

\begin{abstract}
The DESI Collaboration reports a significant preference for a dynamic dark energy model ($w_0w_a$CDM) over the cosmological constant ($\Lambda$CDM) when their data are combined with other frontier cosmological probes. We present a direct Bayesian model comparison using nested sampling to compute the Bayesian evidence, revealing a contrasting conclusion: for the key combination of the DESI DR2 BAO and the Planck CMB data, we find the Bayesian evidence modestly favours $\Lambda$CDM (log-Bayes factor $\ln B = -0.57{\scriptstyle\pm0.26}$), in contrast to the collaboration's 3.1$\sigma$ frequentist significance in favoring $w_0w_a$CDM. Extending this analysis to also combine with the DES-SN5YR supernova catalogue, our Bayesian analysis reaches a significance of $3.07{\scriptstyle\pm0.10}\,\sigma$ in favour of $w_0w_a$CDM. By performing a comprehensive tension analysis, employing five complementary metrics, we pinpoint the origin: a significant ($2.95{\scriptstyle\pm 0.04}\,\sigma$), low-dimensional tension between DESI DR2 and DES-SN5YR that is present only within the $\Lambda$CDM framework. The $w_0w_a$CDM model is preferred precisely because its additional parameters act to resolve this specific dataset conflict. Replacing DES-SN5YR with the recalibrated DES-Dovekie dataset, this tension is reduced and the three-probe Bayesian evidence for $w_0w_a$CDM vanishes ($\ln B = -0.30{\scriptstyle\pm0.19}$). The convergence of our findings with alternative statistical analyses suggests that the preference for dynamic dark energy is primarily driven by the resolution of inter-dataset tensions, warranting a cautious interpretation of its statistical significance.
\end{abstract}

\maketitle

\section{\label{sec:introduction}Introduction}

The DESI DR2 data release~\cite{desi2025} reports up to 4.2$\sigma$ preference for dynamical dark energy ($w_0w_a$CDM) over $\Lambda$CDM based on a frequentist hypothesis test derived from a likelihood ratio based test statistic. This strong evidence comes from analysing the combination of DESI DR2 BAO data, the Planck 2018 CMB measurements~\citep{Planck2018params} and the DES-SN5YR supernova catalogue~\citep{descollaboration2025darkenergysurveycosmology}. This result has generated a great deal of interest in Cosmology, signalling a potential deviation from the standard cosmological model~\citep{CosmoVerseNetwork:2025alb}. Recent independent analyses have questioned these claims from a variety of angles, with particular focus being paid to the robustness of the DES-SN5YR supernova catalogue~\citep{efstathiou_evolving_2025,vincenzi2025comparingdessn5yrpantheonsn} or by examining consistency between DESI data releases~\cite{Efstathiou2025BAO}. The DES-Dovekie recalibration~\citep{Popovic2025DovelkieCalib,Popovic2025Dovekie} subsequently updated the DES~Y5 supernova calibration, combining additional tertiary standard stars with a more flexible calibration model. Despite persistent questions about combining late and early time probes, the community consensus holds that early-time probes alone—combining Planck CMB with DESI BAO data—show a 3$\sigma$ frequentist significance favouring evolving dark energy over $\Lambda$CDM. By established heuristics for interpreting frequentist results, this exceeds the threshold for evidence of new physics~\cite{Lyons:2013yja}.

Bayesian model comparison, well-established in cosmology, offers a complementary approach by calculating the Bayesian evidence—which naturally penalises model complexity through prior volume integration~\cite{2008ConPh..49...71T} (Bayesian Occam's razor). Using both frequentist and Bayesian perspectives has become increasingly common~\citep{Herold:2025hkb}, and for claims as significant as evolving dark energy, employing these established frameworks~\citep{Adame_2025} is essential.

In this letter we present a Bayesian analysis of DESI DR2 using nested sampling with \texttt{PolyChord}~\cite{Handley2015PolychordI,Handley2015PolychordII}. Full details are in our upcoming companion paper~\cite{2026arXiv260305472O}. Here we focus on comparing $\Lambda$CDM and $w_0w_a$CDM for the dataset combinations in DESI's Table VI~\cite{desi2025}. We find that when running this well established Bayesian Model Comparison pipeline, \emph{Bayesian} evidence for evolving dark energy is significantly lower, with the combination of CMB and BAO data favouring the $\Lambda$CDM model. We further show that replacing DES-SN5YR with the recalibrated DES-Dovekie dataset eliminates the Bayesian preference for $w_0w_a$CDM. We conclude with some discussion on the origin of the different conclusions reached by the two methodologies.

\section{\label{sec:methods}Methods}
\begin{table*}
\centering
\begin{tabular}{@{}l@{\hspace{12pt}}r@{\hspace{8pt}}r@{\hspace{16pt}}r@{\hspace{8pt}}r@{}}
\toprule
& \multicolumn{2}{c@{\hspace{16pt}}}{This Work (Bayesian)} & \multicolumn{2}{c}{DESI Collab. (Frequentist)} \\
\cmidrule(lr){2-3} \cmidrule(l){4-5}
Dataset & $\ln B$ & Significance & $\Delta\chi^2_{\mathrm{MAP}}$ & Significance \\
\midrule
\multicolumn{5}{l}{\textit{Individual Datasets}} \\
DESI DR2 & $-1.47{\scriptstyle\pm 0.11}$ & n/a & $-4.7$ & 1.7$\sigma$ \\
DESI DR1 & $-1.64{\scriptstyle\pm 0.10}$ & n/a & --- & --- \\
\midrule
\multicolumn{5}{l}{\textit{Pairwise Combinations}} \\
DESI DR2 + CMB (no lensing) & $-0.38{\scriptstyle\pm 0.25}$ & n/a & $-9.7$ & 2.7$\sigma$ \\
DESI DR1 + CMB (no lensing) & $-0.50{\scriptstyle\pm 0.25}$ & n/a & --- & --- \\
DESI DR2 + CMB & $-0.57{\scriptstyle\pm 0.26}$ & n/a & $-12.5$ & 3.1$\sigma$ \\
DESI DR1 + CMB & $-0.38{\scriptstyle\pm 0.26}$ & n/a & --- & --- \\
DESI DR2 + Pantheon+ & $-2.77{\scriptstyle\pm 0.12}$ & n/a & $-4.9$ & 1.7$\sigma$ \\
DESI DR1 + Pantheon+ & $-2.98{\scriptstyle\pm 0.11}$ & n/a & --- & --- \\
DESI DR2 + Union3 & $+0.25{\scriptstyle\pm 0.12}$ & $1.39{\scriptstyle\pm 0.31}\,\sigma$ & $-10.1$ & 2.7$\sigma$ \\
DESI DR1 + Union3 & $+0.42{\scriptstyle\pm 0.11}$ & $1.59{\scriptstyle\pm 0.10}\,\sigma$ & --- & --- \\
DESI DR2 + DES-SN5YR & $+1.56{\scriptstyle\pm 0.12}$ & $2.33{\scriptstyle\pm 0.06}\,\sigma$ & $-13.6$ & 3.3$\sigma$ \\
DESI DR2 + DES-Dovekie & $-1.63{\scriptstyle\pm 0.12}$ & n/a & --- & --- \\
DESI DR1 + DES-SN5YR & $+0.84{\scriptstyle\pm 0.11}$ & $1.92{\scriptstyle\pm 0.07}\,\sigma$ & --- & --- \\
DESI DR1 + DES-Dovekie & $-1.37{\scriptstyle\pm 0.12}$ & n/a & --- & --- \\
\midrule
\multicolumn{5}{l}{\textit{Triplet Combinations}} \\
DESI DR2 + CMB + Pantheon+ & $-1.70{\scriptstyle\pm 0.26}$ & n/a & $-10.7$ & 2.8$\sigma$ \\
DESI DR2 + CMB + Union3 & $+1.37{\scriptstyle\pm 0.27}$ & $2.23{\scriptstyle\pm 0.15}\,\sigma$ & $-17.4$ & 3.8$\sigma$ \\
DESI DR2 + CMB + DES-SN5YR & $+3.32{\scriptstyle\pm 0.27}$ & $3.07{\scriptstyle\pm 0.10}\,\sigma$ & $-21.0$ & 4.2$\sigma$ \\
DESI DR2 + CMB + DES-Dovekie & $-0.30{\scriptstyle\pm 0.19}$ & n/a & --- & --- \\
\bottomrule
\end{tabular}
\caption{Comparison of Bayesian and frequentist model comparison for $w_0w_a$CDM vs $\Lambda$CDM. DESI results from Table VI of Ref.~\cite{desi2025}. Negative $\ln B$ favours $\Lambda$CDM; negative $\Delta\chi^2_{\mathrm{MAP}}$ favours $w_0w_a$CDM. Bayesian significances are only computed when $\ln B > 0$ (favouring $w_0w_a$CDM); n/a indicates cases where the Bayes factor favours $\Lambda$CDM. DES-Dovekie~\cite{Popovic2025DovelkieCalib,Popovic2025Dovekie} denotes the recalibrated DES Y5 supernova sample.}
\label{tab:comparison}
\end{table*}

We analyse DESI DR2 BAO data~\cite{desi2025,BAOData} combined with Planck 2018 CMB (CamSpec likelihood~\cite{CamSpec2021}) and Type Ia supernovae (Pantheon+~\cite{Scolnic2018}, Union3~\citep{rubin2025unionunitycosmology2000}, DES-SN5YR~\citep{descollaboration2025darkenergysurveycosmology}, DES-Dovekie~\citep{Popovic2025DovelkieCalib,Popovic2025Dovekie}). We use the \texttt{PolyChord}~\cite{Handley2015PolychordI,Handley2015PolychordII} nested sampling algorithm via \texttt{Cobaya}~\cite{Torrado2021Cobaya,cobayaascl} and CAMB~\cite{Lewis:1999bs}. For $w_0w_a$CDM, we adopt DESI's priors ($w_0 \in [-3, 1]$, $w_a \in [-3, 2]$).

We compute Bayesian evidence $\mathcal{Z} = P(D|\mathcal{M})$ and log Bayes factor $\ln B = \ln \mathcal{Z}_{w_0w_a\mathrm{CDM}} - \ln \mathcal{Z}_{\Lambda\mathrm{CDM}}$. Following Trotta~\cite{2008ConPh..49...71T}, we convert Bayes factors to Gaussian significance via (i) Bayes factor to $p$-value using the following relation between an upper bound on the Bayes factor $\bar{B}$ and the $p$-value~\citep{sellke_calibration_2001},
\begin{equation}
B \leq \bar{B} = -\frac{1}{e p \ln p} \quad \text{for } p \le e^{-1},
\end{equation}
and (ii) $p$-value to significance via
\begin{equation}
\sigma = \Phi^{-1}(1-p/2),
\end{equation}
where $\Phi^{-1}$ is the inverse normal cumulative distribution function. Following this inversion we can convert our derived Bayes factors into \emph{upper bounds} on significances for comparison with the frequentist results. Converting from Bayes factors to significances allows direct comparison between frequentist and Bayesian hypothesis testing frameworks, however this should be done with considerable caution~\citep{berger_testing_1987,sellke_calibration_2001, kipping_exoplaneteers_2025}. We have chosen a common and conservative approach to this conversion that ensures we quote a reasonable upper bound on the significance it should be possible to obtain. In a Bayesian setting the best practices for interpreting a Bayes factor are to quote the \emph{betting-odds}, and leave it up to the reader to decide if they would \emph{take the bet}. Despite the subtleties of this conversion and some genuine philosophical differences between the two frameworks, Bayesian model comparison is the more standard approach in the astronomical literature, and any departure from it warrants careful justification. The Bayesian evidence naturally penalises model complexity through prior volume integration, whereas frequentist test statistics evaluate fit quality at a single point in parameter space, and this distinction can lead to opposing conclusions on the preferred model, as we discuss in our companion paper~\cite{2026arXiv260305472O}.

\section{\label{sec:results}Results}

Table~\ref{tab:comparison} compares the DESI Collaboration's frequentist hypothesis tests with our Bayesian model comparison for $w_0w_a$CDM versus $\Lambda$CDM. From a Bayesian perspective, the DESI DR2 data alone penalises the complexity of the $w_0w_a$CDM model, favouring the simpler $\Lambda$CDM with a log-Bayes factor of $\ln B = -1.47{\scriptstyle\pm 0.11}$. This preference for $\Lambda$CDM remains when CMB data are added. The combination of DESI DR2 + CMB, which the DESI Collaboration highlights as a key result, yields $\ln B = -0.57{\scriptstyle\pm 0.26}$ in our analysis. This stands in contrast to the frequentist finding of a 3.1$\sigma$ preference for $w_0w_a$CDM, setting the stage for a methodological comparison. A similar pattern is observed for DESI DR1 data, which also favours $\Lambda$CDM both individually and in combination with the CMB.

A direct comparison between DR1 and DR2 reveals that while the overall Bayesian landscape remains similar, the improved precision of DR2 acts to sharpen existing trends. Individually, both datasets show a consistent, moderate preference for $\Lambda$CDM ($\ln B_{\mathrm{DR2}} = -1.47{\scriptstyle\pm 0.11}$ vs. $\ln B_{\mathrm{DR1}} = -1.64{\scriptstyle\pm 0.10}$). This stability extends to combinations with CMB and Pantheon+ data, where the evidence favouring $\Lambda$CDM changes only marginally between the two releases. The most significant evolution is in combinations with supernova catalogues that prefer dynamical dark energy. In particular, the evidence from the DESI + DES-SN5YR combination is amplified in the new data, with the log-Bayes factor in favour of $w_0w_a$CDM strengthening from $\ln B = +0.84{\scriptstyle\pm 0.11}$ for DR1 to $+1.56{\scriptstyle\pm 0.12}$ for DR2. This suggests that the primary impact of the transition from DR1 to DR2 is not a fundamental shift in the BAO data's model preference, but rather an enhancement of the tensions and synergies observed when combined with external datasets, particularly DES-SN5YR.

\begin{figure*}
\centering
\includegraphics[width=0.48\textwidth]{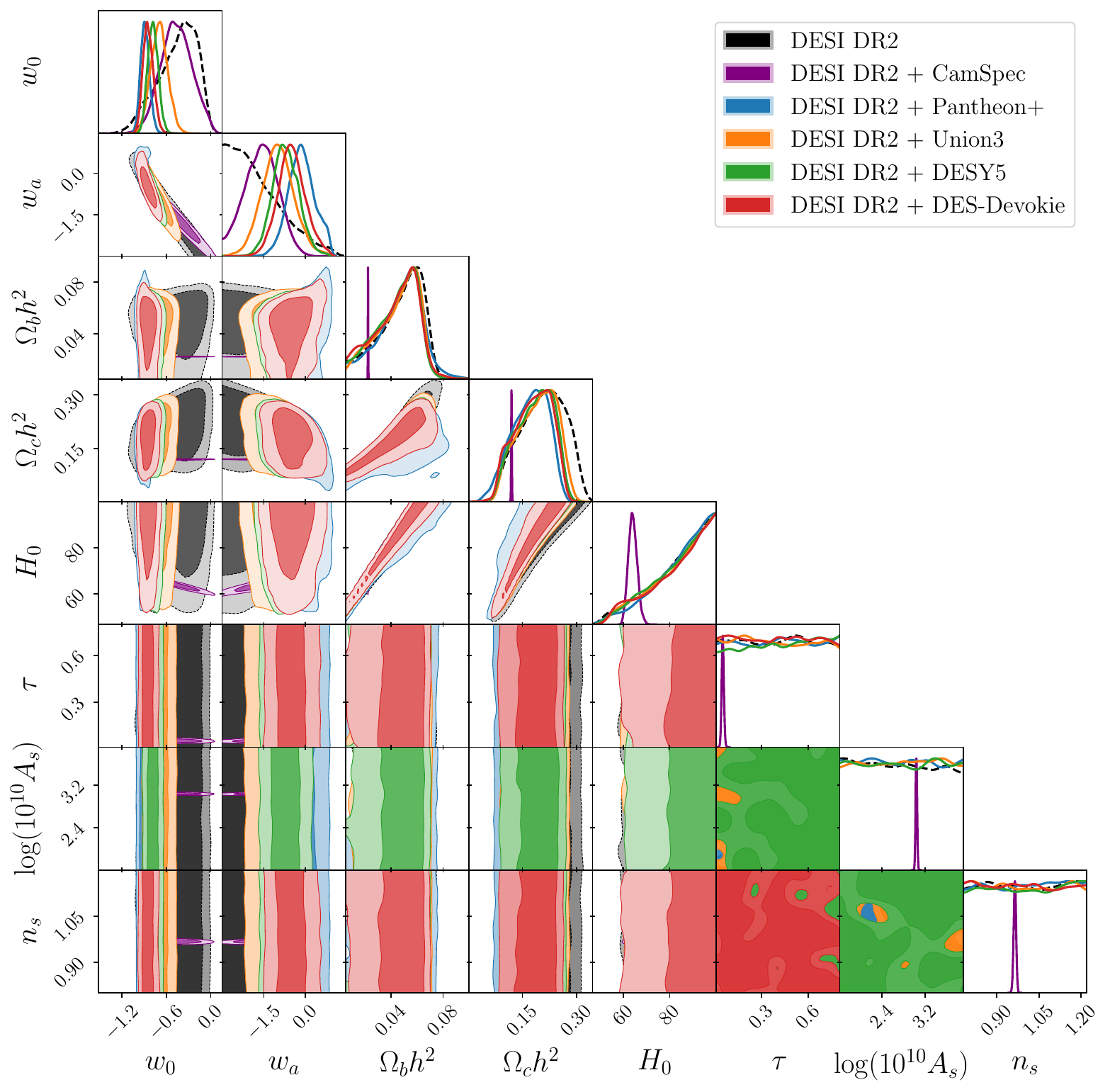}
\hfill
\includegraphics[width=0.48\textwidth]{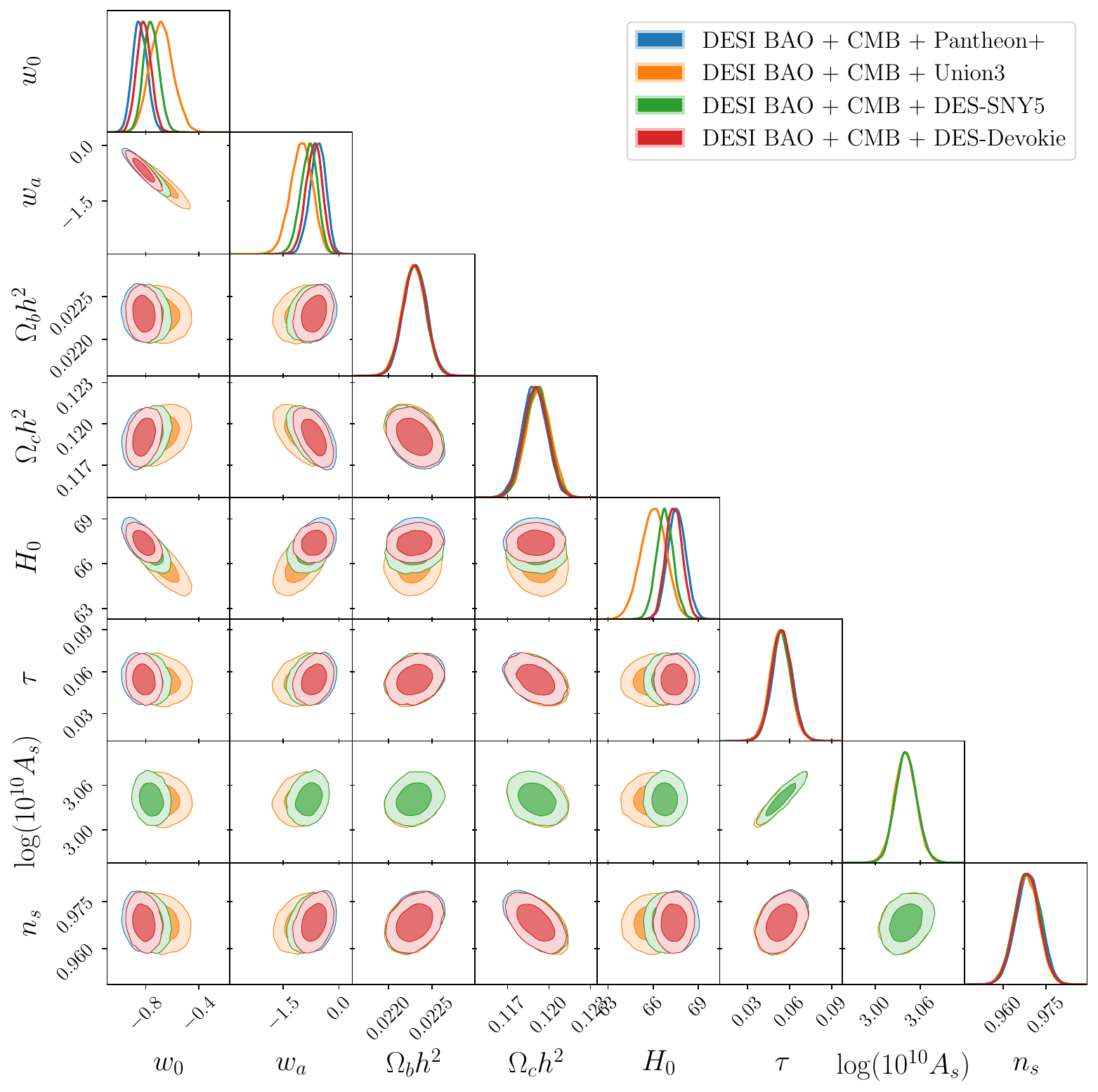}
\caption{Posterior comparisons in $w_0w_a$CDM showing the full cosmological parameter space. \textit{Left:} DESI DR2 alone (black dashed) and pairwise combinations with CMB (purple), Pantheon+ (blue), Union3 (orange), DES-SN5YR (green), and DES-Dovekie (red). \textit{Right:} Triplet combinations with DESI BAO + CMB combined with Pantheon+ (blue), Union3 (orange), DES-SN5YR (green), and DES-Dovekie (red). The differing constraints on $w_0$ and $w_a$ reflect the varying levels of tension between DESI BAO and each additional dataset. Figures produced with \texttt{anesthetic}~\cite{anesthetic}.}
\label{fig:posterior_comparison}
\end{figure*}

The model preference depends critically on the choice of supernova catalogue. Pairwise combinations with Pantheon+ data strengthen the Bayesian evidence for $\Lambda$CDM ($\ln B = -2.77{\scriptstyle\pm 0.12}$). Conversely, the DES-SN5YR catalogue moderately favours $w_0w_a$CDM (pairwise $\ln B = +1.56{\scriptstyle\pm 0.12}$, or $2.33{\scriptstyle\pm 0.06}\,\sigma$). The synergy between probes becomes evident in the triplet combinations: the DESI DR2 + CMB + DES-SN5YR dataset yields strong evidence for dynamical dark energy ($\ln B = +3.32{\scriptstyle\pm 0.27}$, or $3.07{\scriptstyle\pm 0.10}\,\sigma$), reversing the conclusion from DESI+CMB alone. The strong preference for $w_0w_a$CDM emerges exclusively with DES-SN5YR, aligning with ongoing debates about systematic differences between these datasets~\cite{desi2025}. Crucially, even where both methodologies agree on the preferred model (DES-SN5YR combination), the Bayesian significance ($3.07{\scriptstyle\pm 0.10}\,\sigma$) remains substantially weaker than the frequentist claim (4.2$\sigma$). Figure~\ref{fig:posterior_comparison} shows the full cosmological parameter space constraints for these three triplet combinations in $w_0w_a$CDM, illustrating how the choice of supernova catalogue affects not only the dark energy parameters but also the broader cosmological constraints.

To quantify the origin of these differing model preferences, we analyse the statistical consistency between datasets using the suite of metrics provided by the unimpeded evidence framework~\cite{UnimpededPaper,UnimpededSoftware}. For the DESI DR2 + DES-SN5YR pair within the $\Lambda$CDM model, a significant conflict emerges at $\sigma = 2.95 \pm 0.04\sigma$, exceeding our $2.88\sigma$ look-elsewhere threshold. This is the only pairwise combination to yield a negative evidence ratio ($\log R \approx -0.17$), indicating the datasets are mutually inconsistent. A strongly negative suspiciousness ($\log S = -3.83 \pm 0.03$) confirms a direct likelihood conflict, which the Bayesian dimensionality metric diagnoses as a highly localised, low-dimensional conflict ($d_G = 0.989 \pm 0.073$). This specific tension is substantially resolved in $w_0w_a$CDM: $\sigma$ drops to $1.56 \pm 0.03\sigma$, $\log R$ becomes strongly positive ($\approx +3.5$), and the conflict becomes higher-dimensional ($d_G = 3.54 \pm 0.14$), suggesting it is diffused across the larger constrained parameter space. In stark contrast, the DESI DR2 + Pantheon+ pair exhibits only mild tension across all metrics in $\Lambda$CDM ($\sigma = 1.65 \pm 0.03\sigma$, $\log R > 0$) that is not significantly alleviated in the extended model.

This pattern is the same when CMB data are included. The triplet combination of DESI DR2 + CMB + DES-SN5YR sustains a significant tension of $\sigma = 3.00 \pm 0.08\sigma$ in $\Lambda$CDM. The nature of the conflict evolves from a low- to a high-dimensional disagreement ($d_G$ increases from $\approx 1$ to $>3$), and the likelihood-level conflict becomes exceptionally severe ($\log S = -5.56 \pm 0.09$). While the overall evidence ratio remains positive due to the CMB's constraining power, the suite of metrics points to a profound internal inconsistency. Yet, even this more complex tension is reduced in $w_0w_a$CDM, with $\sigma$ dropping to $1.50 \pm 0.05\sigma$ as the conflict is diffused across a larger parameter subspace ($d_G = 5.98 \pm 0.45$). This demonstrates that the strong Bayesian preference for a dynamic dark energy model is driven by its unique capacity to resolve a specific and significant tension introduced by the DES-SN5YR dataset, a tension which persists and becomes more systemic in $\Lambda$CDM when combined with other cosmological probes.

The DES-Dovekie recalibration~\citep{Popovic2025DovelkieCalib,Popovic2025Dovekie} provides a direct test of this diagnosis. Replacing DES-SN5YR with DES-Dovekie, the pairwise tension with DESI DR2 in $\Lambda$CDM drops from $\sigma = 2.95{\scriptstyle\pm 0.04}$ to $1.96{\scriptstyle\pm 0.04}$. Correspondingly, the Bayes factor reverses from $\ln B = +1.56{\scriptstyle\pm 0.12}$ (favouring $w_0w_a$CDM) to $\ln B = -1.63{\scriptstyle\pm 0.12}$ (favouring $\Lambda$CDM). For the three-probe combination DESI DR2 + CMB + DES-Dovekie, we find $\ln B = -0.30{\scriptstyle\pm 0.19}$, showing no Bayesian evidence for dynamical dark energy. These results, broadly consistent with the Bayesian analysis presented in Ref.~\citep{Popovic2025Dovekie}, confirm that the previously reported preference for $w_0w_a$CDM was driven by the DES-SN5YR calibration rather than by a physical signal.

Our finding that DESI DR2 in combination with CMB is consistent with $\Lambda$CDM finds independent support from the alternative statistical analyses of Efstathiou~\cite{Efstathiou2025BAO}. There are well-known pathological regimes where the results of hypothesis tests can differ between Bayesian and frequentist settings~\cite{robert_jeffreys-lindley_2014}, and in these regimes any high-impact claim, such as the discovery of a new physical phenomenon, should be treated with significant caution. We present a detailed analysis of the present discrepancy as an instance of the Jeffreys--Lindley paradox in our companion paper~\cite{2026arXiv260305472O}.

\section{\label{sec:conclusions}Conclusions}

Our Bayesian analysis of DESI BAO reveals a different picture from the frequentist claims of strong evidence for dynamic dark energy. For the key DESI BAO + CMB combination, we find the evidence modestly favours $\Lambda$CDM, in contrast to the reported 3.1$\sigma$ preference for $w_0w_a$CDM. We confirm that the preference for $w_0w_a$CDM is not a feature of the BAO data itself, but emerges exclusively when combined with the DES-SN5YR supernova catalogue. Our tension analysis, using five complementary metrics, pinpoints the origin of this preference: a significant ($2.95{\scriptstyle\pm 0.04}\,\sigma$), low-dimensional tension between DESI BAO and DES-SN5YR within the $\Lambda$CDM framework. The $w_0w_a$CDM model is preferred with this dataset combination precisely because its additional degrees of freedom are effective at resolving this specific conflict, a tension not present with other supernova catalogues like Pantheon+. Replacing DES-SN5YR with the recalibrated DES-Dovekie reduces the inter-dataset tension and eliminates the Bayesian preference for $w_0w_a$CDM ($\ln B = -0.30{\scriptstyle\pm 0.19}$ for the three-probe combination).


Using established cosmological data analysis methodology, we provide a principled benchmark for the \emph{Bayesian} evidence of $w_0w_a$CDM across multiple probes. This exercise reveals a discrepancy between our Bayesian results and the published frequentist hypothesis test results, of a magnitude that cannot reasonably be ascribed to philosophical distinctions between the paradigms. Although both approaches penalize the additional flexibility of $w_0w_a$CDM, they do so via distinct mechanisms. Importantly, our finding that DESI BAO + CMB data are statistically consistent with $\Lambda$CDM is independently corroborated by the alternative statistical analyses of Efstathiou~\cite{Efstathiou2025BAO}, which offers an alternative, principled framework for assessing model consistency. The convergence of Bayesian model comparison, targeted tension diagnostics, and alternative statistical analyses provides a robust statistical basis for the conclusion that claims of evidence for evolving dark energy are overstated and largely driven by tension with a single external dataset. All analysis products are available via the \texttt{unimpeded} framework~\cite{UnimpededPaper,UnimpededSoftware}.

\section{Acknowledgments}
The authors were supported by the research environment and infrastructure of the Handley Lab at the University of Cambridge.
The computations were conducted on DiRAC at the Cambridge Service for Data Driven Discovery (CSD3), operated by the University of Cambridge Research Computing on behalf of the STFC DiRAC HPC Facility (www.dirac.ac.uk). DiRAC is funded by BEIS via STFC capital grants ST/P002307/1 and ST/R002452/1, and operations grant ST/R00689X/1. W.H. is supported by a Royal Society University Research Fellowship.

\bibliography{biblio}

\end{document}